\documentclass{aa}  
\usepackage{lineno,hyperref}
\usepackage{longtable}
\usepackage{graphicx}
\usepackage{natbib}
\usepackage{hyperref}
\bibpunct{(}{)}{;}{a}{}{,} 
\usepackage{url} 
\usepackage{dblfloatfix}  
\usepackage{txfonts}
\usepackage{color}
\begin{document} 

   \title{Classifications for Exoplanet and Exoplanetary Systems – Could it be developed? }

   \subtitle{I.  Exoplanet classification}

   \author{E. Pl\'{a}valov\'a
          \inst{1}
          \and
          A. Rosaev\inst{2}
          }

   \institute{Mathematical Institute, Slovak Academy of Science, Bratislava, Slovakia\\
              \email{plavalova@mat.savba.sk, plavalova@komplet.sk}
         \and
            Research and Educational Center "Nonlinear Dynamics", Yaroslavl State University, Yaroslavl, Russia\\
             \email{hegem@mail.ru}
             }

    \date{Received Month XX, 2024; accepted Month XX, 2024}
    
 
  \abstract
 {When a star is described as a spectral class G2V, we  know that the star is similar to our Sun. We know its approximate mass, temperature, age, and size.  At more than 5,700 exoplanets discovered, it is a natural developmental step to establish a classification for them, such as for example, the Harvard classification for stars. This exoplanet classification has to be easily interpreted and present the most relevant information about exoplanets and divides them into groups based on certain characteristics. } 
   { We propose an exoplanet classification, which using an easily readable code,  may inform you about a planet's main characteristics. The suggested exoplanet classification code contains four parameters by which we can quickly determine the range of temperature, mass, density and eccentricity of exoplanets. }
 {The first parameter concerns the mass of an exoplanet in the form of the units of the mass of other known planets, where e.g.  $M$ represents the mass of Mercury,  $E$ that of Earth,  $N$ Neptune, or $J$ Jupiter. The second parameter is the mean Dyson temperature of the extoplanet's orbit, for which we established four main temperature classes: $F$ represents the $Frozen$ class,  $W$ the $Water$ class,  $G$ the $Gaseous$ class, and $R$ the $Roaster$ class.  The third parameter is eccentricity and the fourth and last parameter is surface attribute which is defined as the bulk density of the exoplanet, where $g$ represents a $gaseous$ planet, $w$- $water$ planet, $t$ - $terrestrial$ planet,   $i$ - $iron$ planet and $s$ - $super$ dense planet.}
{The classification code for Venus, for example, could be $EG0t$ ($E$ - mass in the range of the mass of the Earth, $G$ - $Gaseous$ class, temperature in the range from 450 to 1000 K,  $0$ - circular or nearly circular orbit, $t$ - $terrestrial$ surface), for Earth it could be $EW0t $  ($W$ - $Water$ class - a possible Habitable zone).  }
  {This classification is very helpful in, for example, quickly delimiting if a planet can be found in the Habitable zone; if it is terrestrial or not.  By applying classifications, we can find possible common characteristics of exoplanets, which are often hidden in a miscellaneous database.}


   \maketitle
%
\nolinenumbers
\section{Introduction}
It has been almost thirty years since  the discovery of the first exoplanet 51 Peg b \citep{1995Natur.378..355M}.  Since then,  more than 5 500 exoplanets and at least 900 exoplanetary systems have been were discovered and a number of expected future discoveries is more than optimistic.  For such a number of exoplanets and exoplanetary systems,  it could be useful to have classifications that would help with the search for common characteristics of individual objects and systems.

According to the Cambridge Dictionary\footnote{https://dictionary.cambridge.org/dictionary/english/classification},  classification is "the act or process of dividing things into groups according to their type"and taxonomy is "a system for naming and organizing things, especially plants and animals, into groups that share similar qualities".  One of the most well know classifications/taxonomies are Mendeleev's periodic table of the elements and classification of animals and plants.  These classifications arose due to the necessity to classify individual members based on their general characteristics into certain groups,  classes,  families etc.  Because of the wide-spectral diversity in the exoplanet's properties,  it is necessary to introduce and use a single classification that will allow us to define the common properties of these bodies.

Some of the first suggested taxonomies for exoplanets were proposed by \cite{Sudarsky_2003}, \cite{Marchi_2007} and  \cite{2009A&A...507.1649L}.  These taxonomies have set a benchmark for the spectral division of exoplanets.  These are very precise taxonomies,  but reaching the initial quick data about an exoplanet is very complicated and does not indicate their main features.  In 2012,  at a time when we had identified more than 700 exoplanets, we published the first classification design \citep{2012elbe.workE..19P}.  Since these above mentioned papers,  an other exoplanet classification proposal has been made specifically targeting terrestrial planets \citep{2023MNRAS.519.6210M}.  

The main goal of this paper is to propose a possible classification method for exoplanets,  which can easily and quickly classify a major part of them.  Applying this classification, we are able to classify more than 90 percent of the discovered body of exoplanets.  The rest of them have data pending and can be classified when such data is known.  Another even more important intention of this proposal is to attempt to apply a classification to divide planets into groups,  finding undiscovered dependencies or common characteristics for individual groups of planets.

For practical application we have used the NASA Exoplanet archive\footnote{https://exoplanetarchive.ipac.caltech.edu/},  Planetary Systems Composite Data, which was download 29th July,  2024.

This is the first paper in a series of articles dealing with the classification of exoplanets and planetary systems. The second article in this series focuses on the mass - radius - density diagrams and its dependences, the third on the classification of planetary systems, and the fourth on possible dependencies in the planetary systems.
\section{Planet$/$Exoplanet classification}
The basis of the proposal of this paper of the classification for exoplanets is a taxonomy/classification system described in the paper by \cite{2012elbe.workE..19P}.  In the coming years, with an increasing number of exoplanets discovered and increasing information about individual exoplanets, we have made various changes to the classification and search for a satisfactory form for their classification.  

What are the parameters which are essential to know for characterising an exoplanet?  That is the question.  Perhaps,  the most important are mass, radius or their period? Or is necessary to know eccentricity,  density,  temperature or chemical compound? Or it is necessary to know the composition of the atmosphere or the period of rotation.  All of these parameters can only be very well defined$/$classified for arbitrary exoplanets,  however, on one side contemporary observation devices do not allow us to obtain the values  of many of these mentioned parameters, but on the other side, even if we were to know the values of all of the above parameters,  a classification containing such a large number of parameters would be incomprehensible.  This means that we have to choose three or four of the most important parameters that characterise the exoplanet.  In order for the classification to be effective,  these three or four parameters should be known for a large number of them.

\begin{figure}
   \centering
   \includegraphics[width=\linewidth]{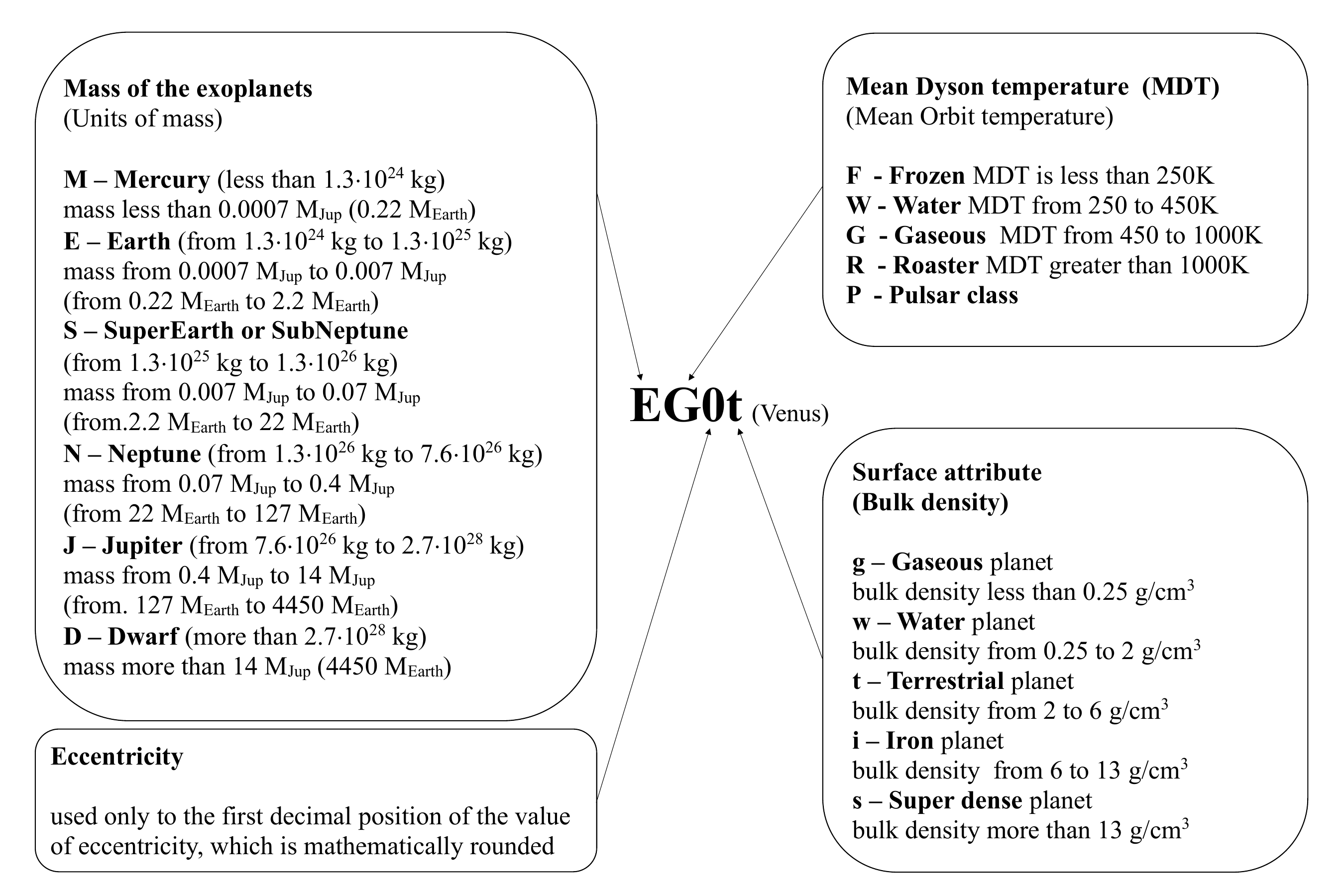}
 
   \caption{Schematic explanation — a presentation of the classification for exoplanets.}
              \label{exoplanet_classif}
    \end{figure}

After lengthy but effective discussions, on workshops,  conferences and other meetings  where a number of positive but also critical opinions were raised, we proposed a revision for the classification of exoplanets.  This classification system incorporates four basic groups; planet mass/mass class,  orbital temperature/MDT, eccentricity and bulk density/surface attribute.  A schematic explanation of our proposal for classification is in Fig. \ref{exoplanet_classif}.
\subsection{The first classification parameter - planet mass}
It is common knowledge that exoplanets with diverse masses often have completely diametrically different properties.  This is an extremely interesting feature that should logically be part of a classification system and for this reason,  the first parameter of the classification concerns the mass of the exoplanet.  

The initial idea \citep{2012elbe.workE..19P} was to divide exoplanets in to mass classes which related to known planets in the Solar System,  with a pattern where the members of each mass class were ten times more massive than members of the previous mass class.  Based on the hypothesis that exoplanets with similar masses likely have the same properties, we revised this empirical (more mathematical) distribution.  After testing many different variants,  in pursuance of actual exoplanet data, with the objective of finding similar behaviour or relations,  we upgraded the previous definitions for mass classes and established the following:

\begin{itemize}
\item
\textbf{M} - \textbf{Mercury class\footnote{Mass of Mercuty is $3.302 \ast 10^{23}  kg$.}} includes exoplanets with a mass less than $0.0007  M_{Jup}$ ($0.22 M_{Earth}$),  meaning exoplanets with a mass less than $1.3 \cdot  10^{24}  kg$.
\item
\textbf{E} - \textbf{Earth class\footnote{Mass of Earth is $5.9736 \ast 10^{24} kg.$}} contains exoplanets with a mass from $0.0007  M_{Jup}$ ($0.22 M_{Earth}$) to $0.007  M_{Jup}$ ($2.2 M_{Earth}$),  meaning exoplanets with a mass in the range from $1.3 \cdot 10^{24}  kg$ to $1.3 \cdot 10^{25}  kg$.
\item
\textbf{S} - \textbf{SuperEarth or SubNeptune class} contains exoplanets with a mass from $0.007  M_{Jup}$ ($2.2 M_{Earth}$) to $0.07  M_{Jup}$ ($22 M_{Earth}$),  meaning exoplanets with a mass in the range from $1.3 \cdot 10^{25}  kg$ to $1.3 \cdot 10^{26}  kg$ .
\item 
\textbf{N} - \textbf{Neputne class\footnote{Mass of Neptune is $1.0243 \ast 10^{26} kg$. }}  contains exoplanets with mass from $0.07  M_{Jup}$ ($22 M_{Earth}$) to $0.4 M_{Jup}$ ($127 M_{Earth}$),  meaning exoplanets with a mass in the range from  $1.3 \cdot 10^{26}  kg$ to $7.6 \cdot 10^{26}  kg$.
\item
\textbf{J} - \textbf{Jupiter class\footnote{Mass of Jupiter is $1.8986 \ast 10^{27} kg$. }} contains exoplanets with mass more than from $0.4  M_{Jup}$ ($127 M_{Earth}$) to $14 M_{Jup}$  ($4 450 M_{Earth}$) meaning exoplanets with a mass in the range from $7.6 \cdot 10^{26}  kg$ to $2.7  \cdot 10^{28}  kg$.
\item
\textbf{D} - \textbf{Dwarf class}  contains exoplanets with mass more than $14 M_{Jup}$ ($4 450 M_{Earth}$)  meaning exoplanets with a mass greater than $2.7  \cdot 10^{28}  kg$.
\end{itemize}

\begin{figure}
   \centering
   \includegraphics[width=\hsize]{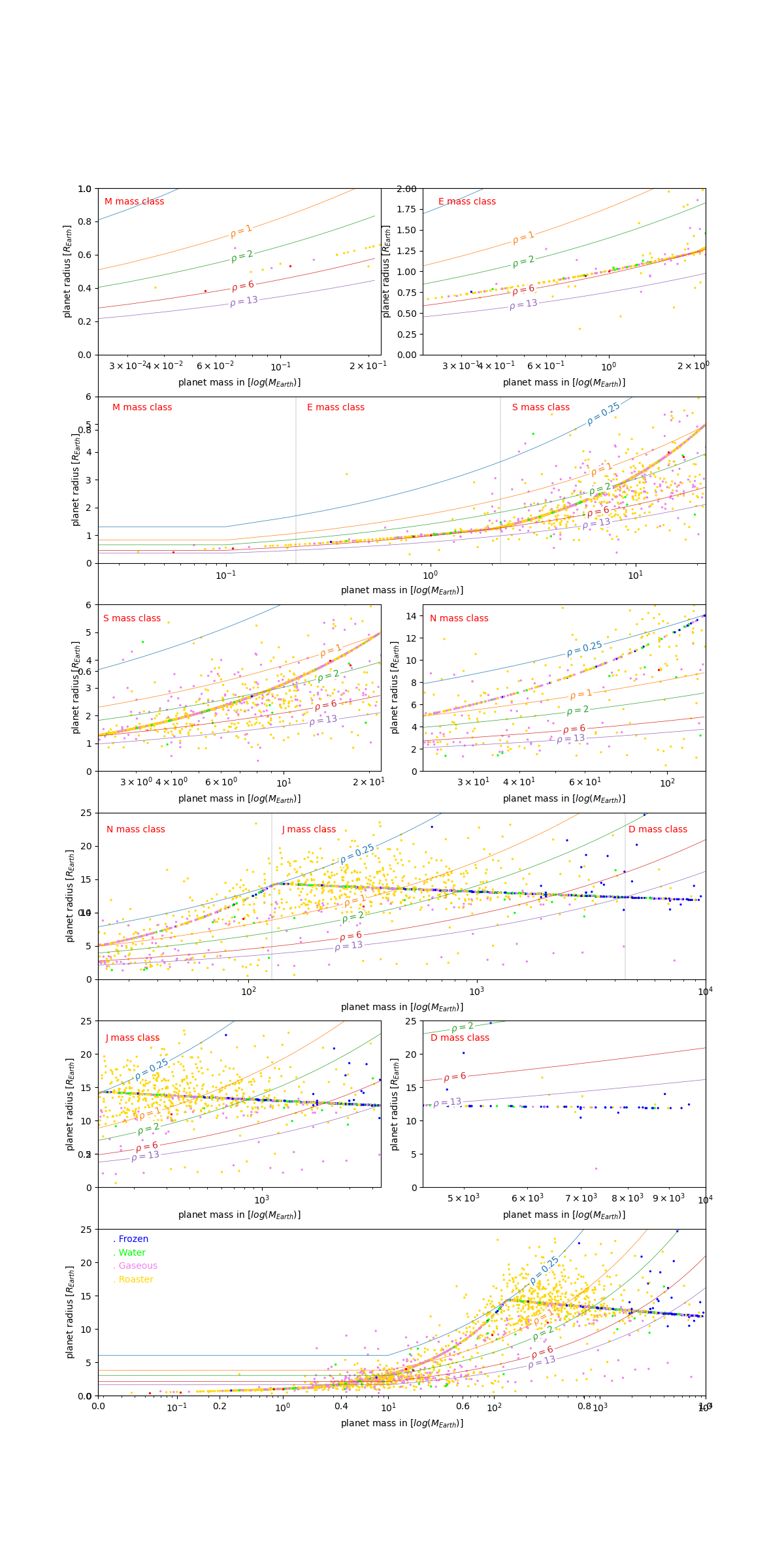}
 
   \caption{M-R diagram according to mass classes.  There are several individual figures for all mass classes and merged figures for some classes which present interesting features.  The colours of the dots are according to the temperature classes.  For comparison, Solar System planets are shown as red dots.}
              \label{mass-radius}
    \end{figure}

In reality,  all the members of both mass classes of $Mercury$ and $Earth$ have a bulk density in range from $ 2 gcm^{-3}$ to  $ 13 gcm^{-3}$ and we could say they are terrestrial planets. We established $Mercury$ class more due to practicality,  but the characteristics of the two groups seem identical.  Due to the higher values of the $Earth$ mass class $0.007  M_{Jup}$ ($2.22 M_{Earth}$), we have placed it in the region where there is a change in the growth rate of the function  (see mass-radius diagram (MR),  Fig \ref{mass-radius},  graph in the second line).

Due to the higher borders for the mass class $S$ - $SuperEarth$ or $SubNeptune $ we set it in an area where the bulk density of the majority of exoplanets reaches $ 1 gcm^{-3}$ (see Fig \ref{mass-radius}, two figures in the third line).

The $Neputne$ mass class contains members with very different values of bulk density.  But we can clearly observe a rise in function  which when reaches a point of $0.4 M_{Jup}$ ($127 M_{Earth}$) immediately changes trajectory (see Fig \ref{mass-radius},  one figure in the fourth line). We have established this point as the upper border for this mass class.  

Close-in planets make up a large part of the $Jupiter$ mass class  (these are the yellow dots ($Roaster$) in the first figure in the fifth line in Fig.  \ref{mass-radius}).  If we do not consider these exoplanets,  the majority of exoplanets in this mass class are grouped into a descending line.  In the area where this descending line intersects the density $ 13 gcm^{-3}$, it is the point of $14 M_{Jup}$  ($4 450 M_{Earth}$),  we decided to define this as the upper limit for the $Jupiter$ mass class.
 
The M-R diagram features several exoplanets,  all of which have very high masses, higher than $14 M_{Jup}$,  and a high value of bulk density,  higher than $ 13 gcm^{-3}$.  According to most catalogues,  these bodies are also exoplanets and therefore we have placed them in  $D$ mass class,  however,  it is very probable that these bodies have different properties, more similar to that of brown dwarfs than exoplanets (hence the idea to name this class $D$ as in brown Dwarf).   In the NASA database used in this paper,  there is an exoplanet named HD 105618c with a mass of $26.5 M_{Jup}$ whereas the mother star Oph 11 has a mass of $0.02 M_{Sun} = 20,96 M_{Jup}$.  While exoplanet Kepler-485b has a mass of $112 M_{Jup}$ and  the mother star TRAPPIST-1 has a mass of $0.09 M_{Sun} = 94.32 M_{Jup}$.

As is clear from the previous paragraph, we used the M-R diagram to define mass classes. The purpose of this article is to clarify the exoplanet classification system, and therefore more detailed analysis of M-R-density dependencies are in the third part of this series.  Although, using actual data,  we tried to find dependencies which could describe the properties of exoplanets,  however, with the help of new future data we could open the discussion for the  revision of defining mass classes.

\subsection{The second classification parameter - temperature MDT}

The value of the surface temperature of an exoplanet depends on many parameters,  for example,  albedo,  speed of the planet rotation or of course of the structure of its atmosphere.  A precise temperature value cannot be determined from observable data.  However, we can define the value of the temperature,  which is in the exoplanet orbit, as an universal independent parameter.

We established a new parameter, the Dyson temperature according to \cite{1960Sci...131.1667D}.  This parameter relates to a temperature that has an artificial sphere the size of a planetary orbit - Dyson sphere. The Dyson temperature can be defined according to the following equation.
\begin{equation}
T_{orb}^4 = \frac{T_{\ast}^4  R_{\ast}^2}{R_{orb}^2} \ ,
\label{Dyson}
\end{equation}
where $R_{orb}$ is the momentary distance from the star to the exoplanet,  $T_{\ast}$ is the effective temperature of the star and $R_{\ast}$ is the radius of the star.  For an exoplanet with a small value of eccentricity,  we could use the results from (\ref{Dyson}) directly.  However,  in the case when an exoplanet has a large value of eccentricity,  the distance from the star changes  according to the Second Keplerian law,  without homogeneity and in many cases rapidly.  Analogically, with high values of the eccentricity,  the values of Dyson temperature also change very rapidly.  These radical changes in Dyson temperature were take into account to implement Mean Dyson Temperature (MDT),  which is the mean value of the calculated Dyson temperature in equal segments of any given orbit according to the Second Keplerian Law (for more information see sect.\ref{MDT}).

We divided the results of MDT into temperature classes mostly according the states of water:
\begin{itemize}
\item
\textbf{F} - \textbf{Frozen class} - where the MDT is less than $250 K$.  The water in this class is in the state of ice.  Jupiter,  Saturn,  Neptune,  and Uranus are members of this class.
\item
\textbf{W} - \textbf{Water class} - where the value of the MDT is in the range from $250 K$ to $450 K$.  Exoplanets in this region could have liquid water in their atmosphere,  surface or inside in their core.  This region we can call the Habitable zone.
\item
\textbf{G} - \textbf{Gaseous class} - where the value of the MDT is in the range from $450 K$ to $1000 K$.  Is very probably that surface water or water found in the core is in gaseous form.
\item
\textbf{R} - \textbf{Roaster class} -  where the value of the MDT is in excess of $1000 K$.  Many of these members are also refer to as hot Jupiters.
\item
\textbf{P} - \textbf{Pulsar class} - as is commonly known,  some exoplanets orbit pulsar stars.  The behaviour of these exoplanets would unquestionably be totally different to exoplanets orbiting main sequence stars and their MDT can not easily be calculated.  However, these are very interesting objects which we  can not exclude from classification  and we have reserved for them the Plusar class.
\end{itemize}
The debate around the edge of the Habitable zone has been a discussion for many years.  With reference to the Solar System, \cite{2017ApJ...837L...4R} determines the inner edge to be $0.95 au$ and the outer edge  $2.44 au$,  whereas according to \cite{2013ApJ...767...65V},  the  borders are from $0.82 au$ to $1.08 au$.   Based on the latest expansions of the boundaries of where life is possible \citep[see,  e.g.,][]{clarke2013low},   for the purpose of the classification of exoplanets,  we decided to extend the borders of the Habitable zone to the maximum range possible.  For the Solar System, in the case when the MDT $=250 K$,  the outer edge of the Habitable zone has a value of $2.485 au$ and when the MDT$=450 K$,  the inner edge of the Habitable zone has a value $0.767 au$.  We are aware that with the growing number of exoplanets discovered and increased knowledge about them,  we will have better clarification of the Habitable zone.  As is evident from the above text that the $Water$ class defines the borders of the Habitable zone. 

The main idea of dividing the temperature range into classes is,  as we have already mentioned,  not only to define the Habitable zone,  but also an effort to identify possible different physical properties of various exoplanet temperature classes.  For this reason we have created two groups of exoplanets with a MDT greater than $450 K$, these being the $Gaseous$ and $Roaster$ classes.  We assume that with a difference of temperature for example a MDT of $700 K$ for one exoplanet and a MDT of $1 800 K$ for another,  vastly different chemical processes or atmospheric conditions or etc.  may occur.  We defined the boundary between the $Gaseous$ and $Roaster$ classes empirically,  and overtime, new data will inevitably clarify the value of these boundaries.

\begin{figure}
   \centering
   \includegraphics[width=10 cm]{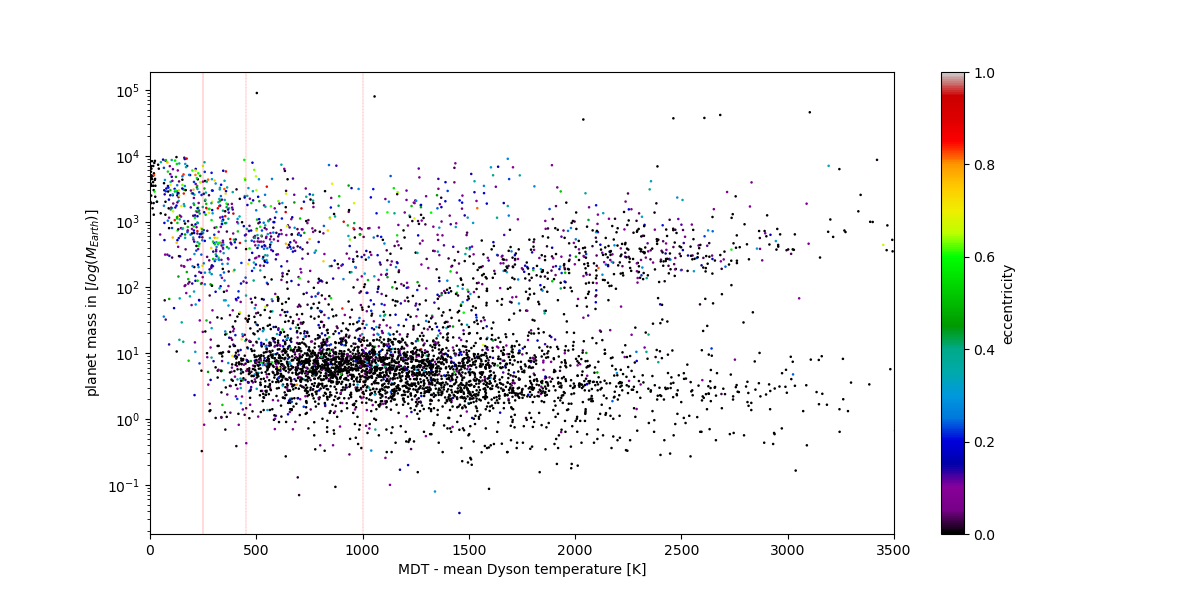}
 
   \caption{The relation between an exoplanet's MDT and its mass.  The dots are coloured  according to eccentricity.}
              \label{MDT-MEarth}
    \end{figure}

In the first data analysis, we did not find a significant dependence between the mass of the exoplanet,  its MDT and eccentricity (see Fig.  \ref{MDT-MEarth}). As mentioned in the previous chapter, a detailed analysis of this data is in a following paper.

\subsection{The third classification parameter - eccentricity}

When an exoplanet has a low value of eccentricity,  its value of Dyson temperature (orbit temperature) remains stable.  However,  when an exoplanet's value of eccentricity is high,  it must be accounted for in our calculations.  Generally,  eccentricity is a very important behavioural characteristic, especially in the discussion around exoplanet climates.  We included eccentricity in the classification as the third parameter.  For easier reference, we have used only the first decimal position for the value of eccentricity,  which is mathematically rounded. For example,  if the value of eccentricity is $0.234$,  we identify it in the classification code as $2$.  For an initial insight in to the range of values expected in an exoplanet's eccentricity,  we have constructed a figure comparing the relation between an exoplanet's MDT and mass,  where the value of eccentricity is colour-coded (Fig.  \ref{MDT-MEarth}).
 \subsection{The fourth classification parameter - bulk density - surface attribute}
How we can describe an exoplanet's surface attributes? One of the tools we can employ to help us understand an exoplanet's surface attributes is transiting spectroscopy \citep[see,  e.g.,][]{2008SSRv..135..345S}.  Despite the fact that from the actual database $74.83\%$ of exoplanlanets have been discovered during transit,  however, transiting spectroscopy is a time-consuming method and also requires high precision instruments.  Other methods such as direct imaging or gravitational microlensing do not provide sufficient data to define the chemical composition or type of exoplanet atmosphere.  

It is difficult, often times impossible, to determine the surface of an exoplanet, but what we can know is its bulk density. From the current data, however, we can obtain the bulk density for most exoplanets,  and for this reason we have introduced this value as the fourth parameter and named it surface attribute.

\begin{figure}
   \centering
   \includegraphics[width=\linewidth]{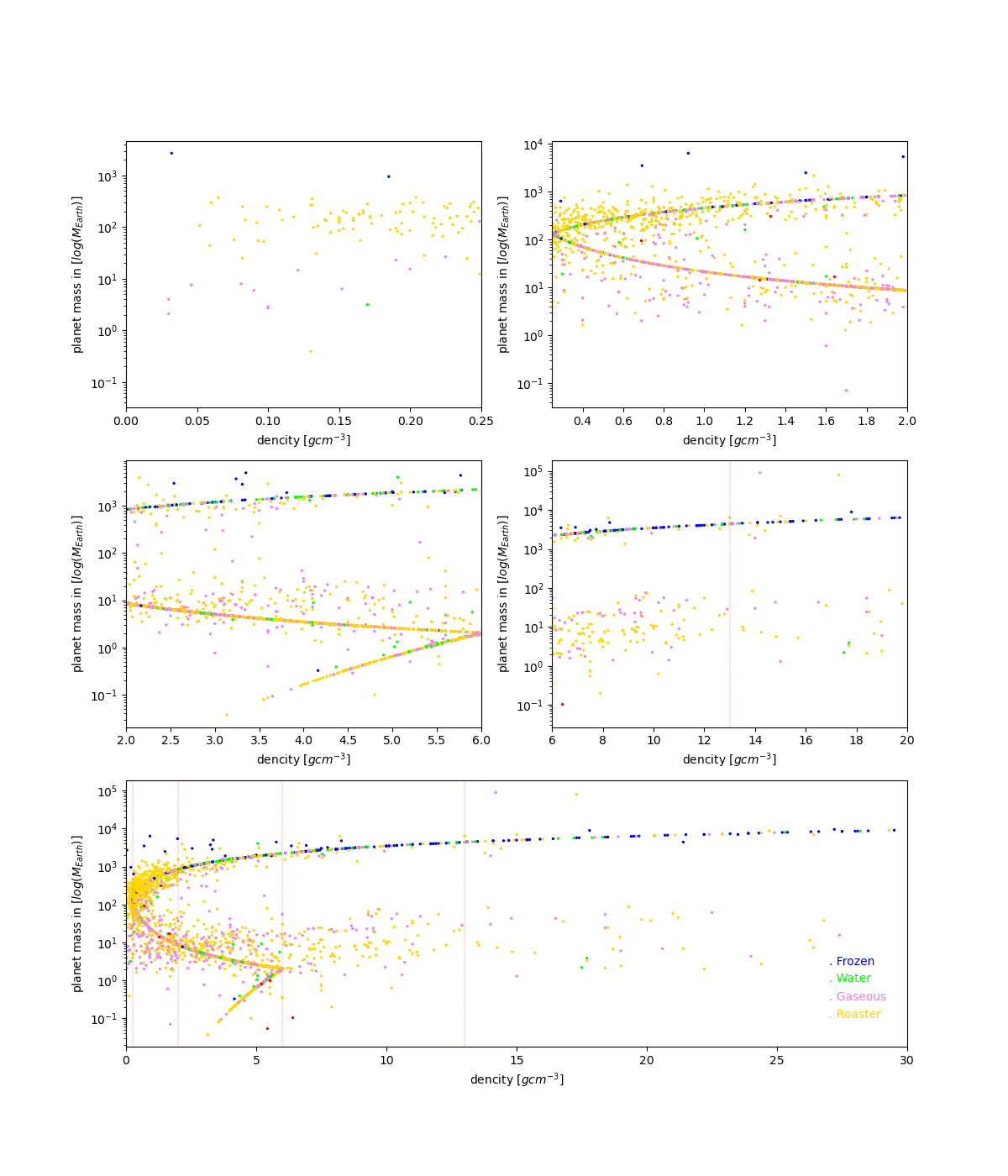}
 
   \caption{The relationship between bulk density and planet mass. The dots are colour-coded according to temperature classes.  The first two lines are divided according to four density classes.  The axes are optimised to better visualise exoplanet properties.  All density classes are shown in the figure in the third line.  For comparison,  Solar System planets are shown as red dots.}
              \label{temperature-radius-dencity}
    \end{figure}

We have introduced these five density groups:
\begin{itemize}
\item
\textbf{g - gaseous} planet contains exoplanets with a bulk density less than $0.25 g/cm^{3}$.  Many of the members of this group are hot close-in exoplanets with a mass around $100 M_{Earth}$ ($\sim 0.3 M_{Jup}$). This group also contains several exoplanets with masses around $10 M_{Earth}$ with a $Gaseous$ MDT.
\item
\textbf{w - water} planet is an exoplanet which has a bulk density is in the range of $0.25 g/cm^{3}$ to $2 g/cm^{3}$.  These exoplanets could have a water/ocean surface.  
\item
\textbf{t - terrestrial} planet is the most interesting group, which contains exoplanets with a bulk density from $2 g/cm^{3}$ to $6 g/cm^{3}$.  As the name suggests, members of this group are terrestrial planets, including Earth, Mars, Venus and Mercury.
\item
\textbf{i - iron} planet is an exoplanet with a bulk density in the range of $6 g/cm^{3}$ to $13 g/cm^{3}$.  Members of this group can be divided into two parts.  Exoplanets with   $Frozen$ or $Water$ (Habitable Zone) MDT  with a mass greater than $10^3 M_{Earth}$ ($\sim 3 M_{Jup}$) is the upper line and the second,  larger group of exoplanets have quite a low mass,  in the order of tens $M_{Earth}$ but with a high MDT in the $Gaseous$ or $Roaster$ class (Fig.  \ref{temperature-radius-dencity} ).
\item
\textbf{s - super-dense} planet is an exoplanet with a bulk density greater to $13 g/cm^{3}$.  The majority of members of this group are exoplanets in the $Frozen$ class with a mass around $500 M_{Earth}$ ($\sim 1,5 M_{Jup}$) or greater.
\end{itemize}

Initially our first proposal was to use the density of water as a benchmark,  but after constructing the relations bulk density to mass figure we made the observation that a value of $0.25 g/cm^{3}$ shows a significant change in the density-mass dependence (see figures in the first line in Fig. \ref{temperature-radius-dencity}).  For this reason we chose that value as one of the boundary values.  We determined another benchmark at the value  of $6 g/cm^{3}$,  based on the bulk density of the terrestrial planets of the Solar system (e.g. bulk density of the Earth is $5.51g/cm^{3}$). We set the value $6 g/cm^{3}$ as the border for this group, because this is another extreme in the density-mass dependence   (see left figure in the second line in Fig. \ref{temperature-radius-dencity}).  

To define the last border $13 g/cm^{3}$ we were inspired by the density of mercury. We have chosen this limit empirically, and it is very likely that the acquisition of data on newly observed exoplanets will shift this limit or this limit will have no justification and will cease to exist. 
 
The main goal of this paper is to explain a system of classification for exoplanets. The mass-radius-density dependence for exoplanets is very interesting and we will analyse it in more detail in the third part of this series of these articles.
 \section{Building classification code}
An effective classification must characterise the basic properties of the investigated element by means of a short code.  We detailed an explanation for each of the four classification parameter in the previous section and now we will show how easy it is to construct a classification code,  e.g. for the Earth.  

The first parameter is planet mass, where for the Earth is $E$ as $Earth$  mass class.

The second parameter is MDT - temperature on planets orbit,  which for the Earth is $392 K$.  This value lies within the boundaries of the $Water$ group (Habitable zone) and this  classification parameter is $W$.

The third parameter is eccentricity.  For Earth the value for eccentricity is $0.017$. According to this scale, the third parameter is $0$.

The fourth and last parameter, the surface attribute for the Earth is logical $t$ as $terrestrial$.

When we combine all four parameters, we get the Earth classification code $EW0t$.  

For Venus, which is sometimes called the "twin" of the Earth, the classification code is $EG0t$. From this code, we immediately see that the planet is terrestrial, but it is a little hotter and hence already outside the Habitable zone.  The Solar System planets with basic data and classification codes are in Tab \ref{Sun}. 

According to planetary systems composite data,  from the NASA Exoplanet archive\footnote{https://exoplanetarchive.ipac.caltech.edu/},  which was downloaded 29-th July 2024 containing a database of 5742 exoplanets (plus 8 Solar system planets).  Applying the proposed classification, we have classified 5204 (plus 8 Solar system planets) of the total number, which is more than 90\% of all known exoplanets. 

 \section{Interpretation examples}
\begin{table*}
\caption{Solar system planets - base data and their classification code.}             
\label{Sun}      
\centering          
\begin{tabular}{c c c r r r r r r r r}     

\hline\hline
  &  & class. & Mass &  Mass & Radius & Radius & Period & a  &  e & density \\   
star  & planet  & code & $[M_{Earth}]$ & $[M_{Jup}]$  &  $[R_{Earth}]$ & $[R_{Jup}]$ & [day] & [au]  &  & $[g/cm^{-3}]$ \\  
\hline 
Sun  &   Uranus  &   SF0w  &   14.52  &   4.57E-2  &   3.98  &   3.63E-1  &   30707.49  &   19.19  &   0.05  &   1.27   \\ 
Sun  &   Venus  &   EG0t  &   0.83  &   2.60E-3  &   0.95  &   8.66E-2  &   224.7  &   0.72  &   0.01  &   5.2   \\ 
Sun  &   Earth  &   EW0t  &   1.0  &   3.15E-3  &   1.0  &   9.11E-2  &   365  &   1  &   0.02  &   5.51   \\ 
Sun  &   Jupiter  &   JF0w  &   317.81  &   1.00E+0  &   10.97  &   1.00E+0  &   4335.35  &   5.2  &   0.05  &   1.33   \\ 
Sun  &   Mercury  &   MG2t  &   0.06  &   1.74E-4  &   0.38  &   3.49E-2  &   87.97  &   0.39  &   0.21  &   5.43   \\ 
Sun  &   Neptun  &   SF0w  &   17.13  &   5.39E-2  &   3.82  &   3.48E-1  &   60224.9  &   30.07  &   0.01  &   1.64   \\ 
Sun  &   Mars  &   MW1i  &   0.11  &   3.40E-4  &   0.53  &   4.86E-2  &   686.96  &   1.52  &   0.09  &   6.39   \\ 
Sun  &   Saturn  &   NF1w  &   95.02  &   2.99E-1  &   9.14  &   8.33E-1  &   10756.2  &   9.54  &   0.05  &   0.69   \\ 
\hline

\end{tabular}
\end{table*}
 How can the classification of exoplanets aid us in our research of them? 
 
Firstly, if we are looking for exoplanets with certain properties, we can very quickly and efficiently filter the desired group in the entire exoplanet database.  For example,  if we are  interested in exoplanets  which are very similar to Earth,  we could set the parameters and filtre from all 5212 classified exoplanets only those  which have the same classification code as Earth ($EW0t$),  revealing ten exoplanets (see Tab \ref{EW0t}). 
According to the needs of the search, we can filtre only one classification parameter (or arbitrary any two or three) depending on what properties we are interested in.

\begin{table*}
\caption{Exoplanets with the same classification code as Earth,  $EW0t$.}             
\label{EW0t}      
\centering          
\begin{tabular}{c c c r r r r r r r r}     

\hline\hline
  &  & class. & Mass &  Mass & Radius & Radius & Period & a  &  e & density \\   
star  & planet  & code & $[M_{Earth}]$ & $[M_{Jup}]$  &  $[R_{Earth}]$ & $[R_{Jup}]$ & [day] & [au]  &  & $[g/cm^{-3}]$ \\  
\hline 
Kepler-186  &   Kepler-186 f  &   EW0t  &   1.71  &   5.38E-3  &   1.17  &   1.07E-1  &   129.94  &   0.43  &   0.04  &   5.87   \\ 
Proxima Cen  &   Proxima Cen b  &   EW0t  &   1.07  &   3.37E-3  &   1.03  &   9.39E-2  &   11.19  &   0.05  &   0.02  &   5.38   \\ 
TOI-700  &   TOI-700 d  &   EW0t  &   1.25  &   3.93E-3  &   1.07  &   9.78E-2  &   37.42  &   0.16  &   0.04  &   5.56   \\ 
TRAPPIST-1  &   TRAPPIST-1 d  &   EW0t  &   0.39  &   1.22E-3  &   0.79  &   7.18E-2  &   4.05  &   0.02  &   0.01  &   4.37   \\ 
TRAPPIST-1  &   TRAPPIST-1 e  &   EW0t  &   0.69  &   2.18E-3  &   0.92  &   8.38E-2  &   6.1  &   0.03  &   0.01  &   4.9   \\ 
TRAPPIST-1  &   TRAPPIST-1 f  &   EW0t  &   1.04  &   3.27E-3  &   1.04  &   9.52E-2  &   9.21  &   0.04  &   0.01  &   5.02   \\ 
TRAPPIST-1  &   TRAPPIST-1 g  &   EW0t  &   1.32  &   4.16E-3  &   1.13  &   1.03E-1  &   12.35  &   0.05  &   0.0  &   5.06   \\ 
Teegarden's Star  &   Teegarden's Star b  &   EW0t  &   1.16  &   3.65E-3  &   1.05  &   9.57E-2  &   4.91  &   0.03  &   0.03  &   5.51   \\ 
Teegarden's Star  &   Teegarden's Star c  &   EW0t  &   1.05  &   3.30E-3  &   1.02  &   9.30E-2  &   11.42  &   0.05  &   0.04  &   5.44   \\ 
Sun  &   Earth  &   EW0t  &   1.0  &   3.15E-3  &   1.0  &   9.11E-2  &   365  &   1  &   0.02  &   5.51   \\ 
\hline

\end{tabular}
\end{table*}
Secondly,  if we have many exoplanets in one system,  by applying this classification to these exoplanets we have a basic picture of the demographic of the system.  For example, we can immediately determine if some exoplanets are located inside the Habitable zone,  and if so,  are these exoplanets terrestrial or not.  We may also sort exoplanets according to mass, semi-major axis or surface attribute.  In Tab \ref{two_systems} is  an example of two exoplanetary systems GJ 3293  and 55 CnC. 

Thirdly,  but we are convinced not lastly,  building relations according to classifications classes and analysing possible dependencies may bring new discoveries in the future. 

------------

\begin{table*}
\caption{Systems 55 Cnc and GJ 3293 - base data and their classification code.}             
\label{two_systems}      
\centering          
\begin{tabular}{c c c r r r r r r r r}     

\hline\hline
  &  & class. & Mass &  Mass & Radius & Radius & Period & a  &  e & density \\   
star  & planet  & code & $[M_{Earth}]$ & $[M_{Jup}]$  &  $[R_{Earth}]$ & $[R_{Jup}]$ & [day] & [au]  &  & $[g/cm^{-3}]$ \\  
\hline 
55 Cnc  &   55 Cnc b  &   JR0w  &   263.98  &   8.31E-1  &   13.9  &   1.27E+0  &   14.65  &   0.11  &   0  &   0.54   \\ 
55 Cnc  &   55 Cnc c  &   NG0w  &   54.47  &   1.71E-1  &   8.51  &   7.76E-1  &   44.4  &   0.24  &   0.03  &   0.49   \\ 
55 Cnc  &   55 Cnc d  &   JF1t  &   1232.49  &   3.88E+0  &   13  &   1.18E+0  &   5574.2  &   5.96  &   0.13  &   3.08   \\ 
55 Cnc  &   55 Cnc e  &   SR0i  &   7.99  &   2.51E-2  &   1.88  &   1.71E-1  &   0.74  &   0.02  &   0.05  &   6.66   \\ 
55 Cnc  &   55 Cnc f  &   NW1w  &   44.81  &   1.41E-1  &   7.59  &   6.92E-1  &   259.88  &   0.77  &   0.08  &   0.56   \\ 
\hline
GJ 3293  &   GJ 3293 b  &   NW1w  &   23.54  &   7.41E-2  &   5.19  &   4.73E-1  &   30.6  &   0.14  &   0.06  &   0.93   \\ 
GJ 3293  &   GJ 3293 c  &   SF1w  &   21.09  &   6.64E-2  &   4.87  &   4.44E-1  &   122.62  &   0.36  &   0.11  &   1   \\ 
GJ 3293  &   GJ 3293 d  &   SW1t  &   7.6  &   2.39E-2  &   2.67  &   2.43E-1  &   48.13  &   0.19  &   0.12  &   2.19   \\ 
GJ 3293  &   GJ 3293 e  &   SG2t  &   3.28  &   1.03E-2  &   1.63  &   1.49E-1  &   13.25  &   0.08  &   0.21  &   4.16   \\ 
\hline

\end{tabular}
\end{table*}
\section{Conclusions}
We have endeavoured to build a classification for exoplanets that,  firstly,  could be used as a quick and easy mechanism to determine their main attributes and,  secondly,  to allow for a quick and clear comparison of large numbers of them. Thirdly,  but not lastly,  discovering new relations according to classifications classes and analysing possible dependencies carries a strong potential to bring new discoveries in the future.  

The suggested exoplanet's classification code contains four parameters:
 \begin{enumerate}
      \item mass -- where there are mass classes: $M-Mercury$ (with a mass less than $1.3 \cdot  10^{24}  kg$),  $E-Earth$ (with a mass in the range from $1.3 \cdot 10^{24}  kg$ to $1.3 \cdot 10^{25}  kg$),  $S-SuperEarth \setminus SubNeptune$ (with a mass in the range from $1.3 \cdot 10^{25}  kg$ to $1.3 \cdot 10^{26}  kg$),  $N-Neptune$ (with a mass in the range from  $1.3 \cdot 10^{26}  kg$ to $7.6 \cdot 10^{26}  kg$),  $J-Jupiter$ (a mass in the range from $7.6 \cdot 10^{26}  kg$ to $2.7  \cdot 10^{28}  kg$) and $D-Dwarf$ (with a mass greater than $2.7  \cdot 10^{28}  kg$).
      \item temperature -- where are temperature classes: $F-Frozen$ (where the MDT is less than $250 K$), $W-Water$, (where the value of the MDT is in the range from $250 K$ to $450 K$),  $G-Gaseous$ (where the value of the MDT is in the range from $450 K$ to $1000 K$) and $R-Roaster$ ( where the value of the MDT is in excess of $1000 K$).
      \item eccentricity -- where in the proposed classification,  we have used only the first decimal position for the value of eccentricity,  which is mathematically rounded.  
      \item bulk density -- where we have placed five density groups: $g-gaseous$ (a bulk density less than $0.25 g/cm^{3}$), $w-water$ (a bulk density in the range from $0.25 g/cm^{3}$ to $2 g/cm^{3}$), $t-terrestrial$ (a bulk density in the range from $2 g/cm^{3}$ to $6 g/cm^{3}$), $i-iron$  (a bulk density in the range from $6 g/cm^{3}$ to $13 g/cm^{3}$) and $s-super-dense$ (a bulk density greater than $13 g/cm^{3}$).
 \end{enumerate}
One of the goals of this classification was to define the $Water$ class as within the borders of the Habitable zone.

According to this classification,  the classification code for Venus, for example, could be $EG0t$ ($E$ - mass in the range of the mass of the Earth, $G$ - $Gaseous$ class, temperature in the range from 450 to 1000 K,  $0$ - circular or nearly circular orbit, $t$ - $terrestrial$ surface), for Earth it could be $EW0t $ ($W$ - $Water$ class - a possible Habitable zone).  Applying the proposed classification, we have classified more than 90\% of all known exoplanets.  The remaining 10\% have insufficient data to be classified.

\begin{acknowledgements}
Part of this research was supported by the Munich Institute for Astro-, Particle and BioPhysics (MIAPbP) which is funded by the Deutsche Forschungsgemeinschaft (DFG, German Research Foundation) under Germany´s Excellence Strategy – EXC-2094 – 390783311.

Special thanks to Joe Minns for proofreading services.


\end{acknowledgements}

%
%
\bibliographystyle{aa}
\bibliography{bibfile_Plavalova.bib}
\section{Additional material}
\subsection{Mean Dyson Temperature (MDT)}
\label{MDT}
Exoplanets with a high value of eccentricity have a varying range of Dyson temperature throughout their orbit.  Such exoplanets could pass through different zones during the course of their orbital period. For example,  exoplanet HD22781 b moves from the $Frozen$ zone to the Habitable zone and then to the $Gaseous$ zone in one orbital period.  Exoplanet HIP 66074 b even passes through all the temperature zones from $Frozen$ to $Roaster$.  The existence of such exoplanets with a high value of the eccentricity brings us to the conclusion that we must take into account these radical change in Dyson temperature (orbital),  and therefore we have  introduced the Mean Dyson Temperature (MDT).

Because the speed of an orbit is not constant, it changes according to the second Keplerian Law and we are not able to accurately calculate the MDT as a mean value from Dyson temperature in its pericentre and apocentre.  We divided the exoplanet's orbit into equal segments according to the time needed for a complete rotation (in this paper we decided on 10 segments).  For this calculation,  we used the Kepler equation.  For the calculation of the eccentric anomaly,  we have used an iteration method with three steps  \citep[see, e.g.,][]{1971znm..book.....A} and calculated the value of the momentary distance of the exoplanet from its star using the following equation \citep[see, e.g.,][] {karttunen2003fundamental}:
\begin{equation}
R_{orb} = a(1-e\cos E)  \ .
\label{momentary_distance}
\end{equation}
Here,  $a$ is the exoplanet's semi-major axis,  $e$ the eccentricity,  and $E$ the eccentric anomaly of the exoplanet.  We then calculated the distance of each of the ten segments applying (\ref{momentary_distance}).  From the calculated distance values,  we have calculated the temperature according (\ref{Dyson}),  which we have named a momentary Dyson temperature.  Applying all the calculated results of the momentary Dyson temperature we calculated the arithmetical mean,  which we defined as the Mean Dyson temperature for the exoplanet's orbit.
\end{document}